\documentclass[doublecol]{epl2}
\usepackage{bm}
\newcommand{\nh}{\hat{n}}
\newcommand{\xv}{{\bf x}}
\newcommand{\rv}{{\bf r}}
\newcommand{\qv}{{\bf q}}

\newcommand{\bfx}{\mathbf{x}}

\newcommand{\be}{\begin{equation}}
\newcommand{\ee}{\end{equation}}
\newcommand{\bea}{\begin{eqnarray}}
\newcommand{\eea}{\end{eqnarray}}

\newcommand{\half}{\frac{1}{2}}

\def\rf#1{(\ref{#1})}

\title{Smectic-glass transition in a liquid crystal cell with a
  ``dirty'' substrate}

\author{Quan Zhang \And Leo Radzihovsky}

\institute{Department of Physics, University of Colorado,
   Boulder, CO 80309, USA}
\abstract{
  We explore the smectic liquid crystal order in a cell with a
  ``dirty'' substrate imposing random pinnings.  Within harmonic
  elasticity we find a subtle three-dimensional surface
  disorder-driven transition into a pinned smectic-glass, controlled
  by a three-dimensional Cardy-Ostlund-like fixed line, akin to a super-rough phase
  of a two-dimensional \textit{xy} model. We compute the associated random
  substrate-driven distortions of the smectic-glass state, identify
  the characteristic length scales on the heterogeneous substrate and
  in the bulk, and discuss a variety of experimental signatures.
}
\begin{document}
\maketitle

\section{Introduction}
Over the past several decades there has been considerable progress in
understanding the phenomenology of ordered condensed matter states
subject to random heterogeneities, generically present in real
materials \cite{FisherPhysicsToday}. More recently, attention turned
to systems where the heterogeneity is confined to a surface
\cite{FeldmanVinokurPRL,usFRGPRL,usFRGPRE}, an important example of
which is a liquid crystal cell with a ``dirty''
substrate \cite{Aryasova, ClarkSmC, CDJonesThesis,NespoulousPRL104}. These
surface-disordered systems are of considerable interest and exhibit
phenomenology qualitatively distinct from their bulk-disordered
counterparts.

The schlieren texture commonly observed in nematic cells~\cite{Schlieren_texture} is a
manifestation of such surface pinning in nematic cells. Recent
experimental studies also include photo-alignment and dynamics in
self-assembled liquid crystalline monolayers~\cite{Fang_photoalignment,
ClarkSAMs}, as well as memory effects and multistability in
the alignment of nematic cells with heterogeneous random-anchoring
substrates~\cite{Aryasova}. The existence of the corresponding phenomena
in smectic liquid crystals has recently been revealed in ferroelectric
smectic-C cells in the book-shelf geometry
\cite{ClarkSmC,CDJonesThesis}. This latter system was found to exhibit
long-scale smectic layer distortions, demonstrated to be driven by
collective random surface-pinning, and awaits a detailed description.
Among many puzzling features observed in this system is the onset of
broadening of the X-ray smectic peak as the temperature is reduced
toward the smectic A-C transition.

\begin{figure}
  \centering
\includegraphics[height=6.3 cm]{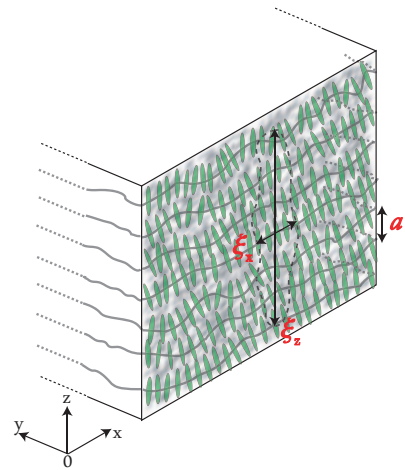}\\      %[height=5 cm]
\caption{(Colour on-line) Schematic of a smectic liquid crystal cell with a
  heterogeneous substrate. The dashed region denotes an anisotropic
  finite-range $\xi_x\times \xi_z$ ($\xi_z\sim \xi_x^2/a$) smectic domain, with
  only short-range smectic order.}
\label{fig:smecticCartoon}
\end{figure}

A schematic of such a smectic liquid-crystal cell is illustrated in
fig.~\ref{fig:smecticCartoon}, with a ``dirty'' front substrate
imposing two types of surface disorders: surface heterogeneity that
pins the nematic director $\hat{n}$ in random planar orientation (random
\textit{orientational} disorder), without a zenithal component, and the surface pinning that imposes a
quenched random potential on the layer position (random \textit{positional}
disorder).  Generically such random surface pinning distorts the smectic
layers and for sufficiently strong pinning can induce topological
defects. As a first treatment of this problem, in this Letter we focus
on the simpler weak-disorder limit, where topological defects are absent
or sufficiently dilute, so that a purely elastic description suffices.

In this Letter we study the correlations of surface disorder-driven
distortion of the smectic layers within a harmonic elastic description. To
summarize our findings, we find that the conventional smectic order is
always unstable in the presence of even infinitesimal random surface
pinning. It is limited to a highly anisotropic finite domain (see
fig.~\ref{fig:smecticCartoon}), characterized by Larkin-like length
scales $\xi_z\sim\xi_x^2/\lambda$ ($z$ along the smectic layer normal),
with $\lambda=\sqrt{K/B}$ a material dependent length set by the bend
and compressional elastic moduli $K$ and $B$ that is (away from the
nematic-smectic transition) typically comparable to the layer thickness
$a$~\cite{deGennesBook}.  On scales smaller than the domain size, the
disorder effects are weak and can be treated perturbatively. In the
plane of the substrate the pinning leads to power-law growth of the
smectic layer distortions and the associated short-range smectic
order, that heal exponentially with the distance (beyond the domain
size) from the substrate into the bulk.

To understand the behavior on longer scales where pinning is dominant,
we employ the functional renormalization group (FRG) to account for
random-potential nonlinearities~\cite{DSFisherFRG}. We find a three-dimensional (3D)
Cardy-Ostlund-like (CO)~\cite{CardyOstlund} phase transition at a
temperature $T_g$, from a weakly disordered smectic for $T>T_g$ (where
on long scales the surface positional pinning is averaged away by
thermal fluctuations), to a low-temperature disorder-dominated
smectic-glass for $T<T_g$. In this phase the positional disorder flows
to a CO fixed line at which correlations of the layer distortions
are described asymptotically by orientational surface pinning alone,
with an effective strength additively enhanced with
$(T_{g}-T)^2$. Namely, long-scale smectic correlations on the substrate are
characterized by $C(x,z)=\overline{\langle [u_0(x,z)-u_0(0,0)]^2
  \rangle}$, with:
\begin{equation}
C(x,z)\sim  \pi a^2\delta_f(T)\left\{\begin{array}{ll}
x/\xi_x^f,& \mbox{ for } x\gg\sqrt{\lambda z},\\
\sqrt{\lambda z}/\xi_x^f,& \mbox{ for } x\ll\sqrt{\lambda z},
\end{array}\right.
\label{large_scale_correlation}
\end{equation}
where the high- and low-temperature phases are distinguished by (among other
features) the effective dimensionless temperature-dependent strength
of the orientational disorder $\delta_f(T)\equiv\Delta_f(T)/\Delta_f$,
given by
\begin{equation}
\delta_f(T)=\left\{\begin{array}{ll}
1,&\mbox{ for }T>T_g,\\
1+
\frac{9}{8\pi^2}\frac{\xi^f_x}{a}\left(1-\frac{T}{T_g}\right)^2,
& \mbox{ for } T\leq T_g.
\end{array}\right.
\label{Delta_f_T}
\end{equation}
The quantity $\xi^f_x$ is given below in eq.~(\ref{LLrandomTilt}).
These predictions remain valid only on scales shorter than the
distance between unbound dislocations and the scale $\xi_{NL}$%$\xi^{NL}_x$
($\gg \xi^f_x$ for weak pinning) beyond which the
nonlinear elasticity may become important \cite{SmecticLongPaper}.

\begin{figure}
\centering
\includegraphics[width=7.5 cm]{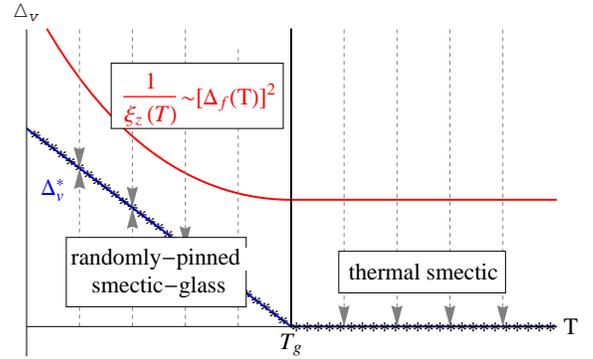}
\caption{(Colour on-line) Temperature-disorder phase diagram for a thick smectic cell
  with a dirty substrate. It illustrates a random-substrate driven
  transition at $T_g$ into a low-temperature pinned smectic-glass
  phase, controlled by a nontrivial fixed line of disorder strength
  $\Delta_v^*(T)\propto (T_g-T)$ for $T<T_g$. The width of the X-ray scattering peak
  $1/\xi_z$, as given in eq.~(\ref{peak_width}), is plotted as the top curve.}
\label{fig:phaseDiagram}
\end{figure}

Based on our finding we suggest that surface disorder is a plausible
explanation for the commonly observed stripes~\cite{CDJonesThesis} in thin
smectic cells~\cite{SmecticLongPaper}. We further argue that this 3D
smectic-glass transition (illustrated in fig.~\ref{fig:phaseDiagram})
may have already been observed as the aforementioned precipitous X-ray
peak broadening in cooled smectic liquid crystal cells with a random
substrate~\cite{CDJonesThesis,ClarkSmC}. Although further detailed systematic
studies are necessary to test this conjecture, based on the
robustness of our theoretical prediction, we expect this transition to
be quite generic in smectic liquid crystal cells with unrubbed
substrates.

\section{Model}
Below we outline the derivation of the above results with the detailed
analysis delayed to a future
publication~\cite{SmecticLongPaper}. Neglecting the elastic
nonlinearities~\cite{SmecticLongPaper,GrinsteinPelcovits}, we model a smectic cell by
the energy functional
\begin{equation}
 H_{bulk}=\int d^{d-1}x\int_0^{\infty} dy
\left[\frac{K}{2}(\nabla^2_\perp u)^2 +\frac{B}{2}(\partial_z  u)^2\right]+H_{pin},
\label{Hsmectic}
\end{equation}
where $u(\xv,y)$ is the distortion of the smectic layers at a point
$\rv=(\xv,y)$, and in 3D, $\xv=(x,z)$ spans a 2D plane parallel to the
substrate.
Even though the nonlinear elasticity is known to be important
in pure \cite{GrinsteinPelcovits} and bulk disordered %ClarkMeyer
smectic systems \cite{RTaerogelPRB}, detailed analysis \cite{SmecticLongPaper}
\footnote{A complete treatment of elastic and pinning
nonlinearities is currently not available.  Our recent analysis \cite{SmecticLongPaper}
shows that on the random substrate the elastic nonlinearities are indeed
relevant (in the RG sense) in 3D, diminishing into the bulk. However, we find that for weak pinning they can only become important on length scales longer than $\xi_{NL} \gg \xi_x$ and can therefore be neglected over a large range of experimentally relevant length scales.}
shows that in this surface disordered
system, it is relevant below lower critical dimensions $d_{lc}^f=4$ for
the random orientational disorder and $d_{lc}^v=6$ for the random positional disorder,
but it is less relevant than the harmonic elasticity at distances
below a length $\xi_{NL}$ much larger than the domain size ($\xi_{NL}\gg \xi_x$). Thus in
the range of interest we can safely ignore the existence of nonlinearity.
The pinning imposed by the surface disorder on the front
($y=0$) substrate is given by~\cite{SmecticLongPaper}
\begin{eqnarray}
H_{pin}&=&-\int d^{d-1}x \left[(\nh\cdot {\bf g}(\xv))^2+U(\xv)\rho(\xv)\right]\nonumber\\
&\approx& -\int d^{d-1}x \bigg[h(\xv)\partial_x u+V(u,\xv)\bigg],
\label{H_pin}
\end{eqnarray}
including the coupling between the nematic director $\hat{n}$ and the local random pinning axis, ${\bf g}(\xv)$, as well as the coupling between the smectic density $\rho(\xv)$ and the local random scalar potential $U(\xv)$. These are determined by the substrate's local chemical and physical structure (composition, roughness, rubbing, etc.).  In the second line of Eq.~(\ref{H_pin}), we approximated this weak pinning by specializing to the smectic state, with the layer normal $\hat{n}$ taken along the $\hat{z} + \delta{{\bf n}}$  and $\delta{{\bf n}} \approx
 -\nabla_{\perp} u$. This leads to $h(\xv) \approx -2g_x(\xv) g_z(\xv)$ and $V(u,\xv)$ a linear functional of the random potential $U(\xv)$ (for more details see Refs.~\cite{SmecticLongPaper,RTaerogelPRB}).
$h(\xv)\partial_x u$ captures the surface random {\em orientational}
pinning of nematogens and smectic layer normals, while $V(u,\xv)$
encodes random pinning of smectic layer {\em positions}, respectively
associated with orientational and positional irregularities of an
unrubbed substrate.  Without loss of generality we take these random
potentials to be characterized by zero-mean Gaussian distributions, with
the variances
\begin{eqnarray}
\overline{ h(\xv)h(\xv')}&=&\Delta_f\delta_a^{d-1}(\xv-\xv'),\\
\overline{ V(u,\xv)V(u',\xv')}&=&R_v(u-u')\delta_a^{d-1}(\xv-\xv'),
\end{eqnarray}
where $\delta_a^{d-1}(\bfx)$ is a short-range function set by the scale on the order of the molecular size. Its precise form has no qualitative affect on the long scale (longer than its range) behavior which is our focus here~\cite{SmecticLongPaper}.

Since the surface disorder is confined to the front substrate at
$y=0$, there are no nonlinearities in the bulk ($y>0$) of the
cell. Consequently, it is convenient to eliminate the bulk
degrees of freedom $u(\xv,y)$ in favor of the distortion on the
heterogeneous substrate, $u_0(\xv)\equiv u(\xv,y=0)$.
For $T=0$, we eliminate $u(\xv,y)$ by solving the Euler-Lagrange
equation $K\nabla_{\perp}^4u-B\partial_z^2u=f(\xv)\delta(y)$, with
$f(\xv)\delta(y)$ representing the boundary condition which imposes
$u_0(\xv)$ on the substrate, and obtain \cite{SmecticLongPaper}
\begin{eqnarray}
&& u(q_x,q_z,y)=u_0(q_x,q_z) e^{-\frac{y}{\sqrt{2\lambda}}\sqrt{\sqrt{\lambda^2q_x^4+q_z^2}+\lambda q_x^2}}\nonumber\\
\hspace{ -2 cm} &&\times\Bigg[\frac{\sqrt{\sqrt{\lambda^2q_x^4+q_z^2}+\lambda q_x^2}}{\sqrt{\sqrt{\lambda^2q_x^4+q_z^2}-\lambda q_x^2}}
 \sin{\left(\frac{y}{\sqrt{2\lambda}}\sqrt{\sqrt{\lambda^2q_x^4+q_z^2}-\lambda
q_x^2}\right)} \nonumber\\
&&
+\cos{\Big(\frac{y}{\sqrt{2\lambda}}\sqrt{\sqrt{\lambda^2q_x^4+q_z^2}-\lambda q_x^2}\Big)}\Bigg].
\label{uqy}
\end{eqnarray}

Substituting (\ref{uqy}) into (\ref{Hsmectic}) and integrating $y$
across a thick cell, $0\leq y<\infty$, the bulk energy reduces to the
surface functional
\begin{eqnarray}
H_{surface}[u_0]&=&\int\frac{d^{d-2}q_xdq_z}{(2\pi)^{d-1}}\half
\Gamma_{\qv}|u_0(q_x,q_z)|^2\nonumber\\
&&-\int d^{d-1}x \bigg[h(\xv)\partial_x u_0+V[u_0,\xv]\bigg],\ \ \
\label{effectiveH0}
\end{eqnarray}
confined to the random substrate at $y=0$, with
\begin{equation}
\Gamma_{\qv}=B\sqrt{2\lambda}\sqrt{\lambda^2q_x^4+q_z^2}
 \sqrt{\sqrt{\lambda^2q_x^4+q_z^2}+\lambda q_x^2}.
 \label{Gamma_q}
\end{equation}

\section{Larkin analysis}
The importance of surface pinning can be assessed by computing the
distortions $\overline{\langle u_0^2\rangle}$ within the Larkin
treatment~\cite{Larkin}, in which a random force (linear)
approximation $F(\xv)=\partial_{u_0}V[u_0(\xv),\xv]\bigg|_{u_0=0}$ is made to
the random potential $V[u_0,\xv]$, with inherited Gaussian statistics
and variance $\overline{F(\xv)F(\xv')}
\equiv\Delta_v\delta_a^{d-1}(\xv-\xv')=-R_v''(0)\delta_a^{d-1}(\xv-\xv')$. Within
this approximation, the correlation of layer distortions on the random
substrate is given by
\begin{equation}
  C_{0}(\qv)=C_T(\qv)+C_{\Delta}(\qv)=\frac{T}{\Gamma_\qv}
  +\frac{\Delta_fq_x^2+\Delta_v}
  {\Gamma_{\qv}^2},\label{C_Larkin}
\end{equation}
where at long scales the first thermal contribution is subdominant to
the random pinning.

The feature of the above linearized (Larkin) approximation is that it
predicts the range of its own validity, limited to short scales where
the distortion $u_0(\xv)$ remains small.  Standard analysis in 3D gives
$\overline{\langle u_0^2(\xv)\rangle} \approx\int
\frac{\Delta_fq_x^2+\Delta_v}{\Gamma_q^2} \frac{dq_xdq_z}{(2\pi)^2}$, that we find to
diverge with substrate extent $(L_x,L_z)$ for surface orientational
disorder, $\Delta_f$, for $d\leq d_{lc}^f=4$, and for surface
positional disorder, $\Delta_v$, for $d\leq d_{lc}^v=6$.  Thus below
these lower critical dimensions (e.g., for a physical 3D cell), an
arbitrarily weak surface disorder destabilizes long-range smectic
order.

We denote the scales at which these distortions grow to the order of
the smectic period $a$ as $\xi_x$ and $\xi_z$ \cite{Larkin},
characterizing the size of the finite, highly anisotropic smectic
domains. In the regimes of dominant orientational pinning we find
%\begin{subequations}
\begin{equation}
\xi_x^f=
c\frac{B^2\lambda^3a^2}{\Delta_f}\approx \sqrt{\lambda\xi_z^f}.
\label{LLrandomTilt}
\end{equation}
For the dominant surface positional disorder we instead find
\begin{equation}
\xi_x^v=
\left(3c\frac{B^2\lambda^3a^2}{\Delta_v}\right)^{1/3}\approx \sqrt{\lambda\xi_z^v},
\label{LLrandomPosition}
\end{equation}
where $c=\frac{4\pi^2}{\pi-2}\approx 34.6$.
Utilizing the relation \rf{uqy} between surface and bulk distortions
indicates that they decay into the bulk exponentially $\sim
e^{-y/\xi_x}$.

\section{Physics beyond domain size}
On length scales longer than the domain size $\xi_{x,z}$, the disorder
imposed distortions $u_0(\xv)$ are large, invalidating the random
force (Larkin) model, and requiring a nonperturbative treatment of
surface random pinning in its fully nonlinear form (\ref{H_pin}). As
with bulk disorder problems, this can be done systematically using an
FRG treatment~\cite{DSFisherFRG,GiamarchiLedoussal,RTaerogelPRB}.

Employing the FRG analysis~\cite{SmecticLongPaper}, we find that for a
physically relevant 3D cell at finite $T$, the problem significantly
simplifies, allowing us to focus on the dominant lowest harmonic
component of the variance function
\begin{equation}
R_v(u)=\Delta_v\cos{(q_0 u)}/q_0^2,
\end{equation}
with $q_0=2\pi/a$, and all higher harmonics irrelevant at long scales
\cite{SmecticLongPaper}.

To treat this leading random pinning nonlinearity we employ the
standard momentum shell RG
\cite{CardyOstlund,Toner_DiVincenzo_sr,SmecticLongPaper}, integrating
out perturbatively (to one-loop) in the surface disorder
nonlinearities the short-scale modes with support in an infinitesimal
momentum shell $ a^{-1} e^{-\delta\ell} < q_x < a^{-1}$.
The result of this coarse-graining is summarized by the flow equations
for the surface disorder strengths,
\begin{eqnarray}
\partial_{\ell}\hat{\Delta}_f&=&
\hat{\Delta}_f+\pi^2\hat{\Delta}_v^2,
\label{Delta_fFlowSimplified}\\
\partial_{\ell}\hat{\Delta}_v&=&(3-\frac{Tq_0^2}{2\pi
  B\lambda})\hat{\Delta}_v-\hat{\Delta}_v^2,
\label{Delta_vFlowSimplified}
\end{eqnarray}
where $\hat{\Delta}_f=8\pi^4 a/\xi_x^f\propto\Delta_f$ and
$\hat{\Delta}_v=6\pi^2(a/\xi_x^v)^3\propto\Delta_v$ are
dimensionless measures of the two types of disorders.

A straightforward analysis of equation (\ref{Delta_vFlowSimplified})
shows that $\hat{\Delta}_v(\ell)$ exhibits two qualitatively distinct
long-scale behaviors.  For $T>T_{g}\equiv 6\pi
B\lambda/q_0^2$ the effective positional pinning
vanishes $\hat{\Delta}_v^*=0$, and for $T<T_{g}$ it flows to a
nontrivial fixed line $\hat{\Delta}_v^*=3(1-T/T_g)$, as illustrated in fig.~\ref{fig:phaseDiagram}.
Then, for large $\ell$ we replace
$\hat{\Delta}_v(\ell)$ inside (\ref{Delta_fFlowSimplified}) by its
fixed point value $\hat{\Delta}_v^*$ and solve the resulting equation
for $\hat{\Delta}_f(\ell)$, finding
\begin{equation}
  \hat{\Delta}_f(\ell)
  =\Big[\hat{\Delta}_f+\pi^2(\hat{\Delta}_v^*)^2\Big]e^{\ell}
  -\pi^2(\hat{\Delta}_v^*)^2.
\label{Delta_f_lowT_solution}
\end{equation}
Although this predicts the same $e^\ell$ asymptotics throughout, the
prefactor depends on the long-scale value of the effective positional
pinning $\hat{\Delta}_v^*(T)$, distinct in the two phases, $\propto
(T_{g}-T)^2$ for $T<T_{g}$ and vanishing for $T > T_g$.

Thus we find that the surface-disordered smectic cell exhibits a 3D
Cardy-Ostlund-like \cite{CardyOstlund} phase transition at $T_{g}=6\pi
B\lambda/q_0^2$, between a high-temperature phase, where at long
scales the surface positional pinning is smoothed away by thermal
fluctuations, controlled by the orientational disorder and
thermal fluctuations, and a low-temperature smectic-glass pinned
phase, controlled by the nontrivial translational-pinning fixed line,
$\hat{\Delta}_v^*(T)$.  This transition to the surface-pinned state is
analogous to the super-rough phase of a crystal surface grown on a
random substrate~\cite{Toner_DiVincenzo_sr}, the vortex glass phase of
flux line vortices confined to a plane in a type II
superconductor~\cite{DSFisherBG,VortexGlass}, and to the 3D smectic
liquid crystals pinned by a random porous environment such as aerogel
\cite{RTaerogelPRB}. Similar to a number of bulk examples (the Cardy-Ostlund model
\cite{CardyOstlund,DSFisherBG,Toner_DiVincenzo_sr,RTaerogelPRB} and the
Kosterlitz-Thouless transitions), here too the phase transition at $T_g$
does not display experimentally observable thermodynamic singularities of
conventional phase transitions. It is thus detected through the change in
the asymptotic behavior of the correlation functions.

To explore the physical distinctions between the two phases, we
compute the smectic correlation function, utilizing the RG matching
method to relate it at long scales (where disorder dominates over
elasticity and must be treated nonperturbatively) to short scales,
where perturbative analysis with effective, length-scale dependent
couplings suffices. We thereby find
\begin{equation}
C(\qv)=\frac{\Delta_f\left[1 + \frac{\xi^f_x}
{8\pi^2a}(\hat{\Delta}_v^*)^2\right]q_x^2}{\Gamma_q^2}
+\frac{{q_x^3a^3}\Delta_v^*}{\Gamma_q^2}.
\label{matching_correlator}
\end{equation}
We note that because of the additional factor of $q_x$ the direct
contribution of the positional disorder (a consequence of the fixed line) is subdominant
at long scales even for $T<T_g$, where $\Delta_v$ is relevant. Thus,
remarkably, at long scales the smectic layer distortions are
controlled by the orientational pinning with the effective
temperature-dependent strength, $\Delta_f(T)\equiv\Delta_f\delta_f(T)$, but with
scaling identical to that of the short-scale linearized model, as
summarized by eqs.~(\ref{large_scale_correlation}), (\ref{Delta_f_T}).

\section{Experimental predictions}
Such an asymptotically growing phonon correlation function translates
into short-range correlated smectic order, with an
anisotropic X-ray scattering peak of widths, $1/\xi_{x,z}$, and growing
as $1/|k_z-q_0|^{3}$ around $q_0$.
The temperature-dependent onset of the enhanced orientational pinning for
$T<T_g$ leads to a nonanalytic dependence of the smectic peak width along $k_z$
\begin{equation}
  1/\xi_z(T)\sim 1/\xi_x^2\sim \delta_f^2(T),
  \label{peak_width}
\end{equation}
as illustrated in fig.~\ref{fig:phaseDiagram}.
We suggest
that this prediction provides a plausible explanation for the
precipitous X-ray peak broadening observed upon cooling of a smectic
liquid-crystal cell with a random
substrate~\cite{CDJonesThesis,ClarkSmC}. Although the compressional
modulus $B(T)$ is also known to decrease on the approach to the
smectic A-C transition\cite{BModulus}, this variation is smooth away
from $T_{AC},T_{NA}$ and can therefore be distinguished from the above
onset behavior at $T_g$. More systematic experimental and theoretical
studies are necessary to explore this further~\cite{SmecticLongPaper}.

Using eq.~(\ref{uqy}), together with a scaling analysis, we predict that
the short-range smectic order persists arbitrarily deep into the bulk,
but with the correlation length at depth $y > \xi_x$ set by $y$
itself. For weak surface pinning smectic layer distortions are extended
beyond the visible wavelength, validating the Mauguin limit, where
linear light polarization ``adiabatically'' follows the local optic
axis.  Under such conditions, transmission polarized light microscopy
directly probes the smectic layer tilt $\partial_x u_0(x,z)$ on the
heterogeneous substrate~\cite{SmecticLongPaper}, similar to a
surface-disordered nematic cell~\cite{usFRGPRE}, and can be used to test our predictions.

\section{Summary}
In summary, we studied the stability and distortions of smectic order
inside a thick liquid crystal cell with a random heterogeneous substrate, but
neglecting the possible role of disorder-induced dislocations and
elastic nonlinearities that we believe is valid for weak
pinning~\cite{SmecticLongPaper}. We found that even arbitrarily weak
random surface pinning reduces smectic order to finite size
domains. By considering the behavior on longer scales we predicted the
existence of a 3D Cardy-Ostlund-like transition at $T_g$ from a weakly
disordered smectic for $T>T_g$ (where on long scales the surface
positional pinning is averaged away by thermal fluctuations), to a
low-temperature disorder-dominated smectic-glass, where pinning is
enhanced below $T_g$. We leave a number of interesting and challenging
questions to a future publication~\cite{SmecticLongPaper}.

\acknowledgments
We thank N. Clark for discussions, J. Maclennan for a careful
reading of the manuscript, and acknowledge support by the NSF
through DMR-1001240 and MRSEC DMR-0820579.

\end{document}